\documentclass{PoS-pdf}

\providecommand{\href}[2]{#2}
\addcontentsline{toc}{section}{References}

\usepackage{amsmath}
\usepackage{slashed}
\usepackage{amssymb}
\usepackage{euscript}
\usepackage{bbm}
\usepackage{bm}
\usepackage[T1]{fontenc}
\usepackage[utf8]{inputenc}
\usepackage{graphicx}

\newcommand{\Ecal}{{\cal E}}
\newcommand{\bk}{{\bf{k}}}
\newcommand{\tbet}{{\tilde{\beta}}}

\title{NLO parton shower for LHC physics - \\hard processes and beyond}

\ShortTitle{NLO parton shower for LHC physics - hard processes and beyond}

\author{\speaker{S. Jadach}%
  \thanks{The partial support
  of the TH Unit of the CERN PH Division for this author is acknowledged.
}\\
   H. Niewodniczański Institute of Nuclear Physics, Polish
   Academy of Sciences, \\ ul. Radzikowskiego 152, 31-342 Krakow, Poland\\
   E-mail: \email{Stanislaw.Jadach@ifj.edu.pl}}

\author{A. Kusina\\
   Southern Methodist University, Dallas, TX 75275, USA\\
   E-mail: \email{akusina@smu.edu}}

\author{M. Skrzypek\\
   H. Niewodniczański Institute of Nuclear Physics, Polish
   Academy of Sciences, \\ ul. Radzikowskiego 152, 31-342 Krakow, Poland\\
   E-mail: \email{Maciej.Skrzypek@ifj.edu.pl}}

\author{M. Slawinska\\
   H. Niewodniczański Institute of Nuclear Physics, Polish
   Academy of Sciences,\\ ul. Radzikowskiego 152, 31-342 Krakow, Poland\\
   E-mail: \email{Magdalena.Slawinska@ifj.edu.pl}}

\abstract{
The new methodology of adding 
QCD NLO corrections in the initial state Monte Carlo parton shower
(hard process part) is presented using
process of the heavy boson production
at the LHC as an example.
Despite the simplified model of the process,
presented numerical results prove that
the basic concept of the new methodology
works correctly in the numerical environment of the Monte Carlo parton shower
event generator.
The presented method is an alternative to the well established methods,
MC@NLO and POWHEG.
Refinements of the new method with better computer CPU time
efficiency are also discussed.

}

\FullConference{“Loops and Legs in Quantum Field Theory ” 11th DESY Workshop on Elementary Particle Physics,\\
                 April 15-20, 2012,
                 Wernigerode, Germany}

\begin{document}

\section{Introduction}
The Large Hadron Collider (LHC) at CERN
provides rich harvest of experimental data.
The proper understanding and interpretation of these data,
possibly leading to discovery of new phenomena,
requires perfect mastering of the ``trivial'' effects due to the multiple
emissions of soft and collinear gluons and quarks.
Perturbative Quantum Chromodynamics
(pQCD)~\cite{GWP,Gross:1974cs,Georgi:1951sr},
supplemented with clever modelling of the low energy
nonperturbative effects, is an indispensable tool
for disentangling the Standard Model physics component in the data.
This work presents part of the global effort of improving quality
of the pQCD calculations for LHC experiments.

Most of the results presented here are described 
in refs.~\cite{Jadach:2011cr} and \cite{Jadach:2012vs}.
Although this work elaborates on the improved method of
the pQCD calculation combining NLO-corrected hard process
and LO parton shower Monte Carlo (MC), it should be regarded
as the first step towards NNLO-corrected hard process combined
with the NLO parton shower MC~\cite{IFJPAN-IV-2012-7}.

\section{Basic LO parton shower MC}
The multigluon distribution of the single initial state
ladder, which is a building block of our parton shower MC,
is represented by the integrand of the ``exclusive/unintegrated PDF'',
which in the LO approximation is the following:
\begin{equation}
\label{eq:LOMC}
\begin{split}
& 
D(t,x)
=\int dx_0\; dZ\; 
 \delta_{x=x_0 Z}\;
d_0(\hat{t}_0,x_0)\; G(t, \hat{t}_0- \ln x_0 | Z),
\\&
G(t, t_0 | Z)=
 e^{-S_F} \sum_{n=0}^\infty
\bigg(
\prod_{i=1}^n
 \int d^3\Ecal(\bar{k}_i)\;
 \theta_{\xi_i>\xi_{i-1}}
 \frac{2C_F\alpha_s}{\pi^2} \bar{P}(z_i)
\bigg)
\\&
\qquad\qquad\quad\times
\theta_{t>\xi_n}
\delta_{Z=\prod_{j=1}^n z_j},
\end{split}
\end{equation}
where  evolution kernel is $\bar{P}(z)=\frac{1}{2}(1+z^2)$,
evolution time is $\hat{t}_0=\ln(q_0/\Lambda)$ and
the ``eikonal''  phase space integration element is
$
d^3\Ecal(k)=\frac{d^3 k}{2k^0}\;\frac{1}{\bk^2}
            =\pi \frac{d\phi}{2\pi} \frac{d k^+}{k^+} d \xi
$
and $k^\pm = k^0\pm k^3$.
We use rapidities $\xi_i =\frac{1}{2}\ln\frac{k^-_i}{k^+_i}\big|_{\rm Rh}$ 
in the hadron beam rest frame (Rh), and
$\eta_i=\frac{1}{2}\ln\frac{k^+_i}{k^-_i}\big|_{\rm RFHP}$
defined in hard process rest frame (RFHP). 
They are related by $\xi_i=\ln\frac{\sqrt{s}}{m_h}-\eta_i$.
Rapidity  ordering is now 
$t=\xi_{\max}>\xi_n>\dots>\xi_i>\xi_{i-1}>\dots>\xi_0=t_0$,
where $t_0=\xi_0=\ln(q_0/m_h)-\ln x_0$.
The direction of the $z$ axis in the RFHP is 
pointing out towards the hadron momentum.
A lightcone variable of the emitted gluon is defined
as $\alpha_i= \frac{2k_i^+}{\sqrt{s}}$ and
of the emitter parton (quark) as
$x_i=x_0-\sum_{j=0}^{i}\; \alpha_j$ (after $i$ emissions).
We also use fractions $z_i=x_i/x_{i-1}$.
The Sudakov formfactor $S_F$ comes from the ``unitarity'' condition%
\footnote{
   The usual cutoff $1-z<\epsilon$ regularizing the IR singularity
   is implicit.}
$
\int_0^1 dZ\; G(t, t_0 | Z)=1,
$
which is also instrumental in the Markovian MC implementation
used to obtain $D(t,x)$ at any value of $t>t_0$.

The initial distribution 
$d_0(q_0,x_0)$ related to experiment,
to previous steps in the MC ladder,
or to PDF in the standard $\overline{MS}$ system
is not essential for the following discussion,
we only note that the unitarity condition provides
baryon number conservation sum rule
$\int_0^1 dx\; D(t,x) = \int_0^1 dx_0\; d_0(t_0,x_0)$.

For testing our new method of correcting hard process to the NLO level
we use the following simplified
MC parton shower, implementing the DY process with
two ladders and the hard process:%
   \footnote{Following ref.~\cite{Jadach:2011cr},
   we adopt $d\tau_2(P;q_1,q_2) =
   \delta^{(4)}(P-q_1-q_2)\frac{d^3q_1}{2q_1^0}\frac{d^3q_2}{2q_2^0}$.}
\begin{equation}
\label{eq:LOMCFBmaster}
\begin{split}
&\sigma_0=
\int d x_{0F} d x_{0B}\;\;
  d_0(\hat{t}_0,x_{0F}) d_0(\hat{t}_0,x_{0B}) 
 \sum_{n_1=0}^\infty\;
 \sum_{n_2=0}^\infty
 \int dx _F\; dx_B\;
\\&~~~~~~~~~\times
e^{-S_{_F}}
\int_{\Xi<\eta_{n_1}}
\bigg(
\prod_{i=1}^{n_1}
 d^3\Ecal(\bar{k}_i)
 \theta_{\eta_i<\eta_{i-1}}
 \frac{2C_F\alpha_s}{\pi^2} \bar{P}(z_{Fi})
\bigg)
\delta_{x_F = x_{0F}\prod_{i=1}^{n_1} z_{Fi}}
\\&~~~~~~~~~\times
e^{-S_{_B}}
\int_{\Xi>\eta_{n_2}}
\bigg(
\prod_{j=1}^{n_2}
 d^3\Ecal(\bar{k}_j)
 \theta_{\eta_j>\eta_{j-1}}
 \frac{2C_F\alpha_s}{\pi^2} \bar{P}(z_{Bj})
\bigg)
\delta_{x_B = x_{0B}\prod_{j=1}^{n_2} z_{Bj}}
\\&~~~~~~~~~\times
 d\tau_2(P-\sum_{j=1}^{n_1+n_2} k_j ;q_1,q_2)\;
\frac{d\sigma_B}{d\Omega}(sx_Fx_B,\hat\theta)\;
W^{NLO}_{MC}.
\end{split}
\end{equation}
In the LO approximation $W^{NLO}_{MC}=1$.
Rapidity $\xi$ is translated into $\eta$
-- the center of mass system rapidity,
in the forward part (F) of the phase space as
$\xi_i=\ln\frac{\sqrt{s}}{m_h}-\eta_i$,
$\eta_{0F}>\eta_i>\Xi$,
and in the backward (B) part
as $\xi_i=-\ln\frac{\sqrt{s}}{m_h}+\eta_i$, 
$\Xi>\eta_i>\eta_{0B}$.
The rapidity boundary between the two hemispheres $\Xi=0$ is used,
until a more sophisticated version related to
rapidity of the produced $W/Z$ is introduced.

Analytical integration of eq.~(\ref{eq:LOMCFBmaster})
results in the standard factorization formula ($W^{NLO}_{MC}=1$)
\begin{equation}
\label{eq:LOfactDY2LO}
\sigma_0 =
\int_0^1 dx_F\;dx_B\;
D_F(t, x_F)\; D_B(t, x_B)\;
\sigma_B(sx_Fx_B).
\end{equation}
The distributions
$D_F(t, x_F)=(d_0\otimes G_F) (t, x_F)$ and 
$D_B(t, x_B)=(d_0\otimes G_B) (t, x_B)$ are
obtained from separate Markovian LO Monte Carlo runs.
The above LO formula is exact,
and can be tested with an arbitrary numerical precision.

\begin{figure}
  \centering
  {\includegraphics[width=0.8\textwidth,height=65mm]{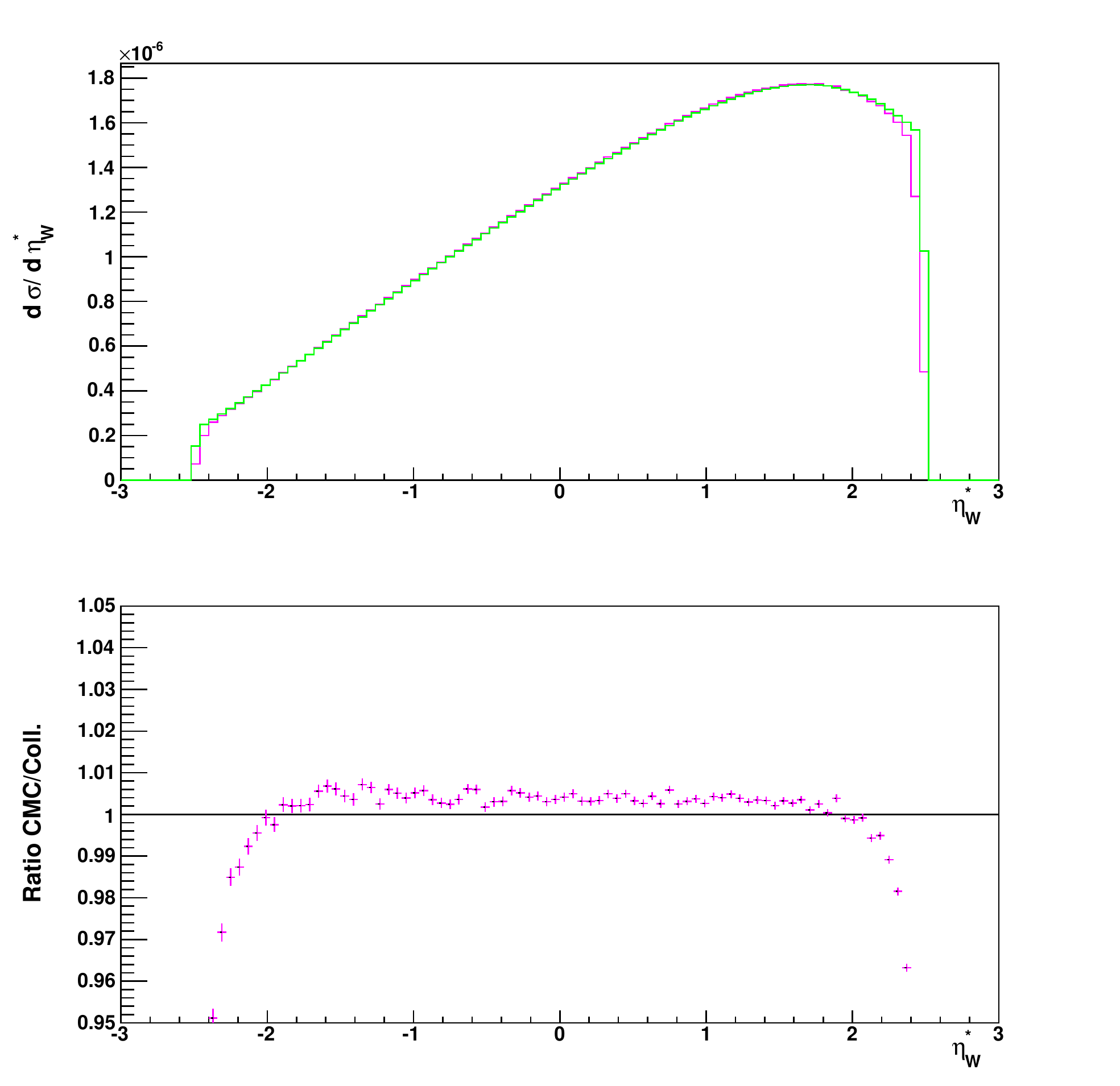}}
  \caption{
    The upper plot shows
    the LO distribution of $\eta_W^*=\frac{1}{2}\ln(x_F/x_B)$ from the CMC
    LO parton shower (purple) and from the strictly collinear formula (green).
    The lower plot shows the ratio of the two.
    }
  \label{fig:etaW_LO_7TeV}
\end{figure}
Figure~\ref{fig:etaW_LO_7TeV} 
represents a ``calibration benchmark'' for the overall normalization
at the LO level.
We show there
the properly normalized distribution of the variable
$\eta_W^*=\frac{1}{2}\ln(x_F/x_B)$, which
in the collinear limit approximates the rapidity of $W$ boson.
The distribution in the upper plot of Fig.~\ref{fig:etaW_LO_7TeV},
representing eq.~(\ref{eq:LOfactDY2LO}),
is obtained using the general purpose MC program FOAM~\cite{foam:2002}.
The collinear PDF $D(t,x)$ there
has been obtained from a separate high statistics run
($10^{10}$ events) of a Markovian MC (MMC),
creating $D(t,x)$ in a form of the 2-dimensional look-up table%
\footnote{
  This MMC run solves the LO DGLAP equation 
  using the MC method,
  as in refs.~\cite{Jadach:2008nu,GolecBiernat:2006xw}.}.
The other distribution in the upper plot of Fig.~\ref{fig:etaW_LO_7TeV} 
represents eq.~(\ref{eq:LOMCFBmaster}) in LO approximation.
It comes from the full scale MC generation 
(with four-momenta conservation).
The MC run with $10^8$ events was used.
The constrained MC (CMC) technique of ref.~\cite{Jadach:2007qa}
is used here because
of the narrow Breit-Wigner peak due to a heavy boson propagator%
\footnote{A backward evolution algorithm of 
ref.~\cite{Sjostrand:1985xi} could be also used here.}.
Two CMC modules and FOAM are combined into one MC generating
gluon emissions and the $W$ boson production.
FOAM is taking care of the generation of the variables
$x_F,x_B,x_{F0},x_{B0}$ and the sharp Breit-Wigner peak
in $\hat{s}=s x_F x_B$, then two CMC
modules are initialized and generate the gluon four-momenta $\bar{k}^\mu_j$.
They are mapped into $k^\mu_j$, following the prescription
defined in ref.~\cite{Jadach:2011cr}, such that
the overall energy-momentum conservation is achieved.
Figure~\ref{fig:etaW_LO_7TeV} demonstrates
a very good numerical agreement
between $d\sigma/ d \eta_W^*$ 
from our full scale LO parton shower MC
of eq.~(\ref{eq:LOMCFBmaster})
and the simple formula of eq.~(\ref{eq:LOfactDY2LO}),
to within 0.5\%,
as seen from the ratio of the two results in the lower part of the figure.

\section{Introducing NLO corrections to hard process}

The NLO corrections to hard process are imposed 
on top of the LO distributions of eq.~(\ref{eq:LOMCFBmaster})
using a single ``monolithic'' weight $W^{NLO}_{MC}$
defined exactly as in ref.~\cite{Jadach:2011cr}:
\begin{equation}
\label{eq:NLODYMCwt}
\begin{split}
&W^{NLO}_{MC}=
1+\Delta_{S+V}
+\sum_{j\in F} 
 \frac{\tbet_1(q_1,q_2,\bar{k}_j)}%
      {\bar{P}(z_{Fj})\;d\sigma_B(\hat{s},\hat\theta)/d\Omega}
+\sum_{j\in B} 
 \frac{\tbet_1(q_1,q_2,\bar{k}_j)}%
      {\bar{P}(z_{Bj})\;d\sigma_B(\hat{s},\hat\theta)/d\Omega},
\end{split}
\end{equation}
the NLO soft+virtual correction is
$
\Delta_{V+S}
=\frac{C_F \alpha_s}{\pi}\; \left( \frac{2}{3}\pi^2 -\frac{5}{4} \right)
$,
and the real correction reads:
\begin{equation}
\label{eq:DYbeta1FB}
\begin{split}
&\tbet_1(q_1,q_2,k)=
\Big[
  \frac{(1-\beta)^2}{2}
  \frac{d\sigma_{B}}{d\Omega_q}(\hat{s},\theta_{F})
 +\frac{(1-\alpha)^2}{2}
  \frac{d\sigma_{B}}{d\Omega_q}(\hat{s},\theta_{B})
\Big]
\\&~~~~~~~~~~~~~~~~~~~~
-\theta_{\alpha>\beta}
 \frac{1+(1-\alpha-\beta)^2}{2}
 \frac{d\sigma_{B}}{d\Omega_q}(\hat{s},\hat\theta)
-\theta_{\alpha<\beta}
 \frac{1+(1-\alpha-\beta)^2}{2}
 \frac{d\sigma_{B}}{d\Omega_q}(\hat{s},\hat\theta).
\end{split}
\end{equation}
The above is the exact ME
of the quark-antiquark annihilation into a heavy vector boson
with additional single real gluon emission%
\footnote{
 We employ here the compact
 representation of ref.~\cite{Berends:1980jk},
 which has also been used in POWHEG~\cite{Alioli:2011nr}.
}.
The LO component, which is already included in the LO MC, is subtracted here.
The variable $\hat{s}=s x_F x_B = (q_1+q_2)^2$
is the effective mass squared of the heavy vector boson.
The definition of angle $\hat\theta$ in the LO component is rather arbitrary.
We define it in the rest frame of the heavy boson,
where $\vec{q}_1+\vec{q}_2=0$,
as an angle between the decay lepton momentum $\vec{q}_1$
and the difference of momenta of the incoming quark and antiquark
$\hat\theta=\angle(\vec{q}_1,\vec{p}_{0F}-\vec{p}_{0B})$.
On the other hand the two angles in the NLO ME are defined 
quite unambiguously as
$\hat\theta_F=\angle(\vec{q}_1,-\vec{p}_{0B})$ and
$\hat\theta_B=\angle(\vec{q}_1, \vec{p}_{0F})$.
In the above we only need directions of the
$\vec{p}_{0F}$ and $\vec{p}_{0B}$ vectors,
which are the same as the directions of the hadron beams.
The lightcone variables $\alpha_j$ and $\beta_j$
of the emitted gluon
are defined in the F and B parts of the phase space as follows%
\footnote{See ref.~\cite{Jadach:2011cr} for more explanations.
}:
\[
\begin{split}
&\alpha_j=1-z_{Fj},\quad
\beta_j=  \alpha_j\; e^{2(\eta_j-\Xi)},\quad ~~
{\rm for}~~~ j\in F,
\\&
\beta_j=1-z_{Bj},\quad
\alpha_j=\beta_j\;   e^{-2(\eta_j-\Xi)},\quad
{\rm for}~~~ j\in B.
\end{split}
\]

Again, the exact phase space integration of eq.~(\ref{eq:LOMCFBmaster}) 
including $W^{NLO}_{MC}$ of eq.~(\ref{eq:NLODYMCwt})
is feasible, and
the resulting compact expression for the
total cross section is obtained~\cite{Jadach:2011cr}:
\begin{equation}
\label{eq:DYanxch}
\begin{split}
\sigma_1 &=
\int_0^1 dx_F\;dx_B\; dz\;
D_F(t, x_F)\;D_B(t, x_B)\;
\sigma_B(szx_Fx_B)
\big\{
\delta_{z=1}(1+\Delta_{S+V})
+C_{2r}(z)
\big\},
\end{split}
\end{equation}
where
$
C_{2r}(z) =\frac{2C_F \alpha_s}{\pi}\; \left[ -\frac{1}{2}(1-z) \right].
$

\begin{figure}[!t]
  \centering
  {\includegraphics[width=0.8\textwidth,height=70mm]{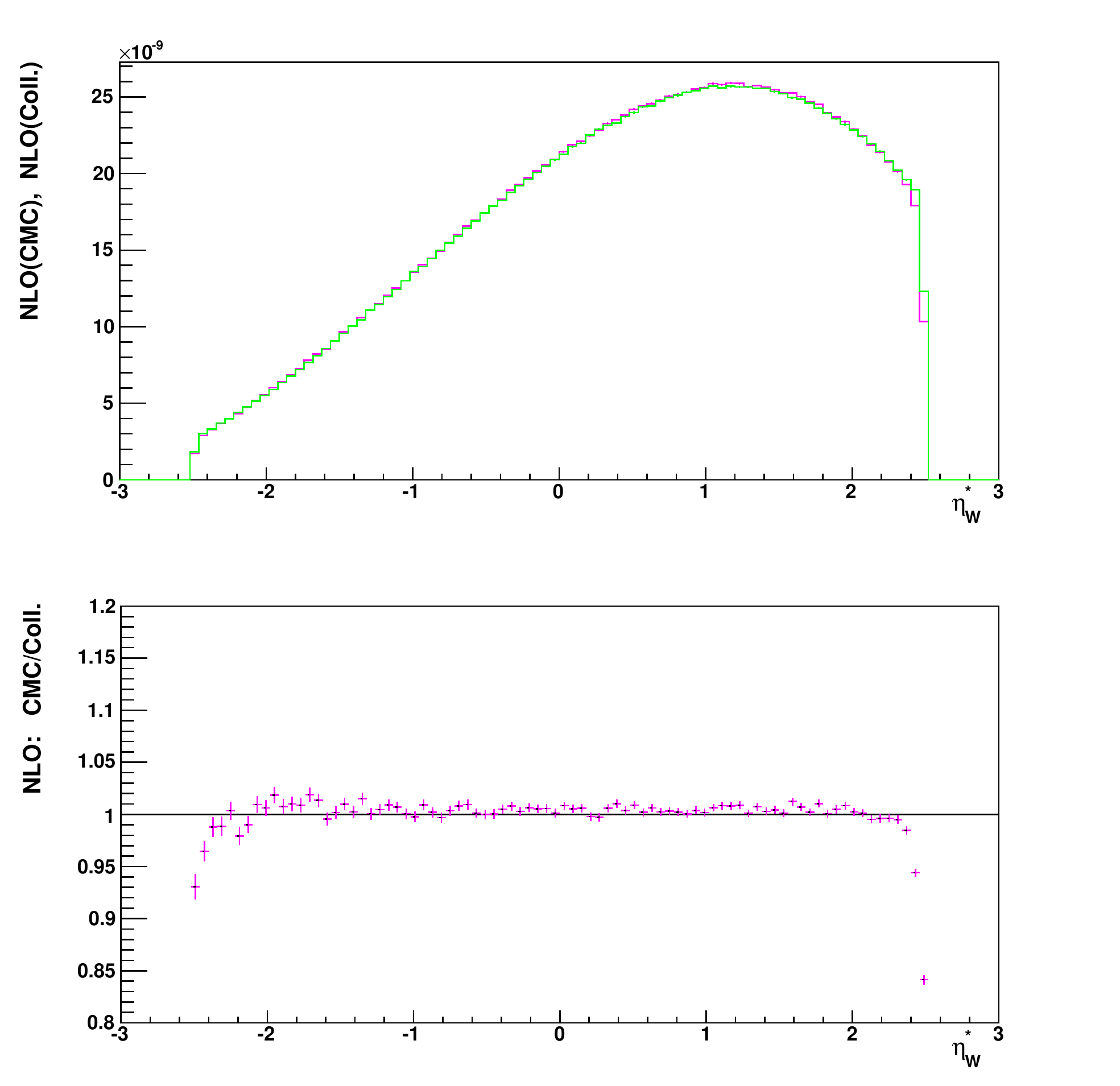}}
  \caption{
    The upper plot shows $(-)$~NLO correction (alone) to the distribution of 
    $\eta_W^*=\frac{1}{2}\ln(x_F/x_B)$ from
    the parton shower (purple)
    and from the the strictly collinear formula (green).
    Their ratio is plotted below.
    }
  \label{fig:etaW_NLOglu_7TeV}
\end{figure}

\subsection{Numerical test of NLO correction}

Figure~\ref{fig:etaW_NLOglu_7TeV} represents a principal 
{\em proof of concept} of our new
methodology for implementing the NLO corrections to the hard process
in the parton shower MC.
The plotted NLO correction to the $\eta_W^*$ distribution%
\footnote{Extra minus sign introduced to facilitate visualization.}
comes from the parton shower MC with the NLO-corrected hard process
according to eqs.~(\ref{eq:LOMCFBmaster}) and~(\ref{eq:NLODYMCwt}).
Additionally we also plot there result
of a simple collinear formula of eq.~(\ref{eq:DYanxch}),
where two collinear PDFs are convoluted with
the analytical coefficient function $C_{2r}(z)$ for the hard process.
Both results coincide within the
statistical error, see their ratio in the lower part of
Fig.~\ref{fig:etaW_NLOglu_7TeV}.

\begin{figure}[!t]
  \centering
  {\includegraphics[width=0.6\textwidth,height=40mm]{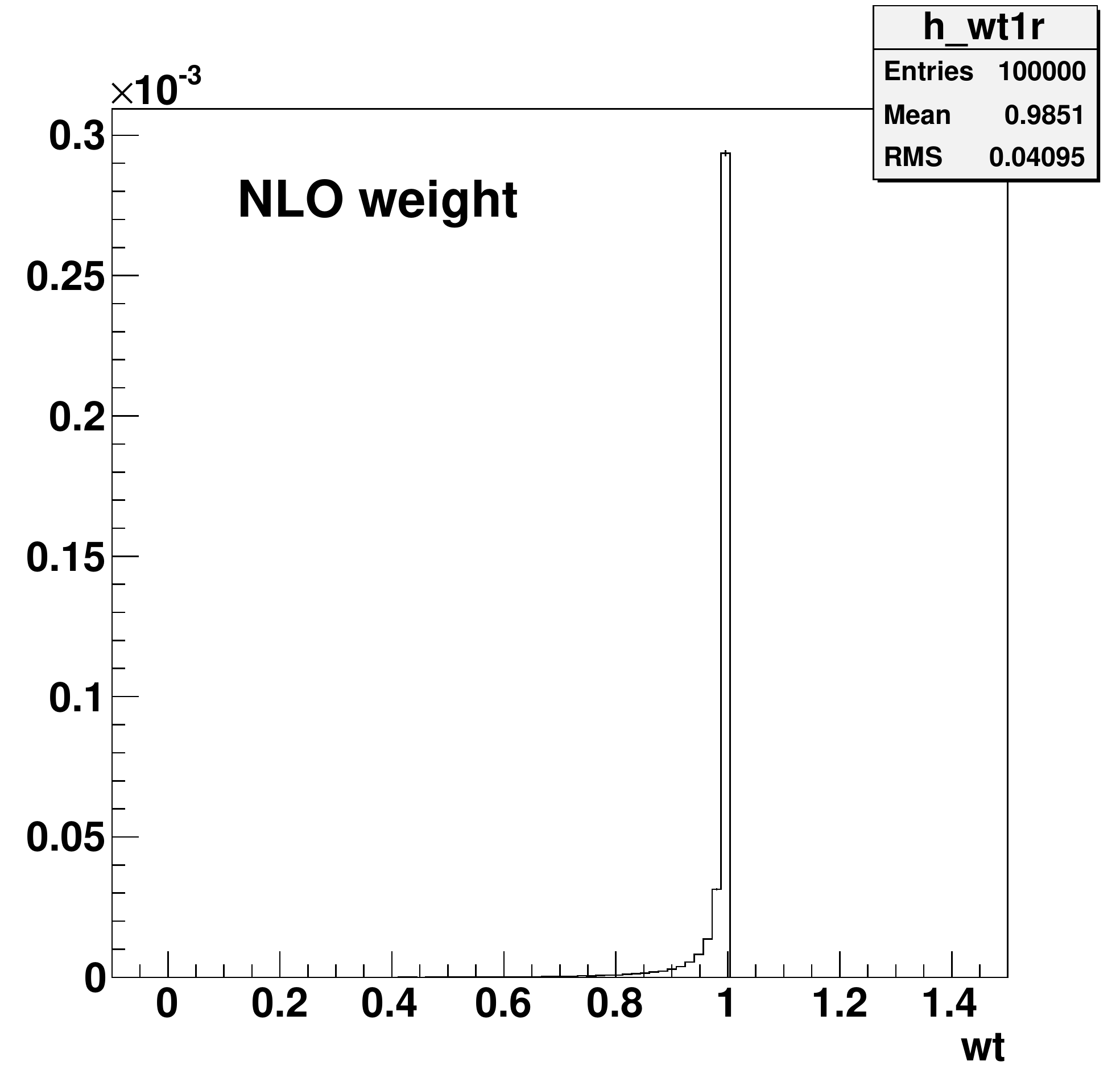}}
  \caption{
    The distribution of the NLO weight 
    $W^{NLO}_{MC}$ of eq.~(\protect\ref{eq:NLODYMCwt}).
    }
  \label{fig:Canv1w_7TeV}
\end{figure}

Technically, the inclusion of the NLO correction
in our parton shower MC is rather straightforward,
and is obtained by including $W^{NLO}_{MC}$ weight of eq.~(\ref{eq:NLODYMCwt}).
MC is providing both LO and NLO-corrected results in a single run
with weighted events.
The NLO weight is strongly peaked near
$W^{NLO}_{MC}=1$, positive, and without long-range tails.
Its distribution is shown in Fig.~\ref{fig:Canv1w_7TeV}.

In all numerical results we have set
$\Delta_{V+S}=0$, as it is completely unimportant
for the presented analysis.
The initial distributions $d_0(q_0,x_0)$ are defined in ref.~\cite{Jadach:2012vs}.

\section{Simplification of the method and comparison with other methodologies}

\begin{figure}[!ht]
  \begin{centering}
  \includegraphics[width=75mm]{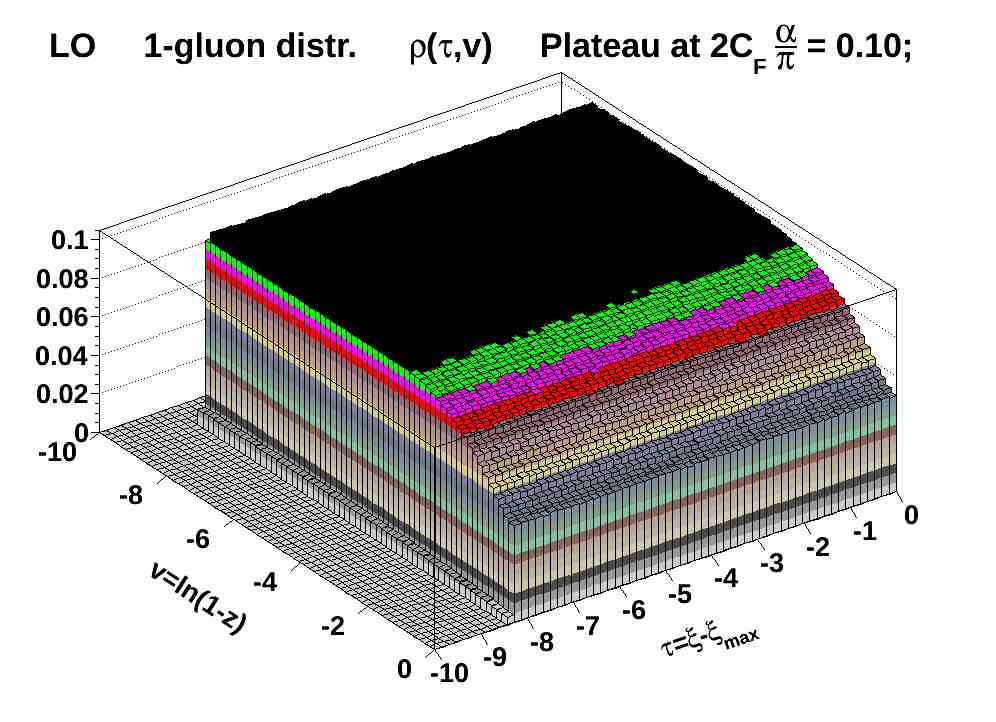}
  \includegraphics[width=75mm]{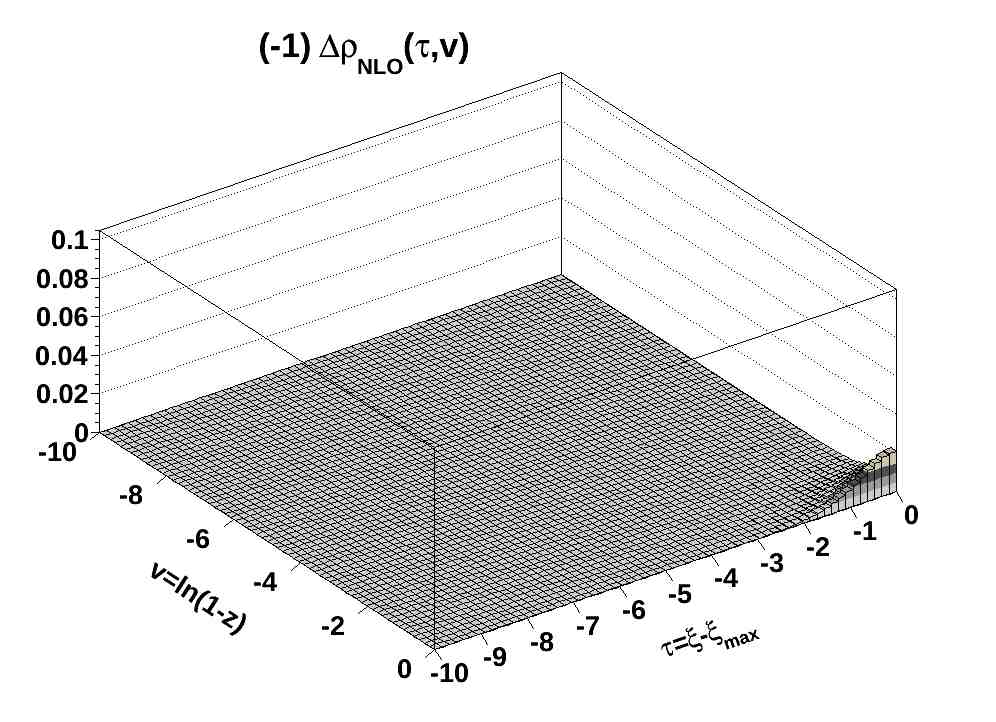}
  \label{fig:mcRho1glu}
  \caption{
    The inclusive distribution of gluons on the log Sudakov plane
    of rapidity $t=\xi$ and $v=\ln(1-z)$ (LHS).
    Contributions from all gluons weighted
    with the component weight $W^{NLO}_j$ (RHS).
    }
  \end{centering}
\end{figure}

Our new method for introducing NLO corrections in
the hard process, proposed in ref.~\cite{Jadach:2011cr}
and tested in ref.~\cite{Jadach:2012vs}, is an alternative
to the two well established
MC@NLO~\cite{Frixione:2002ik}
and POWHEG~\cite{Nason:2004rx,Frixione:2007vw}
methodologies.
With MC numerical implementation at hand, let us elaborate
on the differences with the above two techniques
in particular with the POWHEG technique.
We shall also see that it is possible to make our
method more efficient in terms of CPU time consumption.
This improvement is not so critical in the present case
of NLO corrected hard process, but may be quite useful
in the case of correcting evolution kernels to the NLO
in the ladder parts of the MC~\cite{IFJPAN-IV-2012-7}.

The most important differences 
with the POWHEG and MC@NLO techniques are:
\begin{itemize}
\item
The summation over all emitted gluons,
without deciding which gluon is the one involved in the NLO correction
and which ones are merely ``LO spectators'' in the parton shower.
\item
The absence of $(1/(1-z))_+$ distributions 
in the real part of the NLO corrections 
(virtual+soft correction is kinematically independent).
\end{itemize}
To explain more clearly how
$W^{NLO}_{MC}$ of eq.~(\ref{eq:NLODYMCwt})
is distributed over the multigluon phase space,
we restrict now to single ladder (hemisphere)
with a simplified weight:
\begin{equation}
\label{eq:NLODYMCwt_sim}
\begin{split}
&W^{NLO}_{MC}= 1+\sum_{j\in F} W^{NLO}_j,\qquad
W^{NLO}_j=
 \frac{\tbet_1(q_1,q_2,\bar{k}_j)}%
      {\bar{P}(z_{Fj})\;d\sigma_B(\hat{s},\hat\theta)/d\Omega}.
\end{split}
\end{equation}
In order to find out the phase space regions specific for NLO corrections
we consider inclusive distributions of gluons
on the Sudakov logarithmic plane of rapidity $\xi$
and variable $v=\ln(1-z)$.
In the left hand side (LHS) of Fig.~\ref{fig:mcRho1glu}
we show gluons inclusive distribution in the LO approximation.
The flat plateau there represents IR/collinear singularity%
\footnote{We use constant $\alpha_S$.}
$2C_F\frac{\alpha_S}{\pi} d\xi\frac{dz}{1-z}$
with the drop by factor 1/2 towards $z=0$,
due to $\frac{1+z^2}{2}$ factor in the LO kernel.
In the right hand side (RHS) of Fig.~\ref{fig:mcRho1glu}
we show contributions from all gluons weighted
with the component weight%
\footnote{
 We again insert a minus sign in order to facilitate visualization.}
$-W^{NLO}_j$ of eq.~(\ref{eq:NLODYMCwt_sim}).
The NLO contribution is concentrated in
the area near the hard process rapidity $t=\xi_{\max}$,
which has to be true for the genuine NLO contribution%
   \footnote{It also vanishes towards the soft limit $z\to 1$.}.
The completeness of the phase space near this
important region ($z=0$, $\xi_{\max}$) is critical
for the completeness of the NLO corrections.
Both POWHEG and MC@NLO use standard LO MCs which
feature an empty ``dead zone'' in this phase space corner.

Figure~\ref{fig:mcRho1glu} suggests that
the dominant contribution to $\sum_j W^{NLO}_j$ could be
from the gluon with the maximum $\ln k_j^T\sim \xi_j+\ln(1-z_j)$,
which is closest to the hard process phase space corner.
In the MC we may easily relabel generated gluons using new index $K$
such that they are ordered in the variable 
$\kappa_K=\xi_K+\ln(1-z_K),\; \kappa_{K+1}<\kappa_K$
with $K=1$ being the hardest one.

\begin{figure*}[!ht]
  \centering
  {\includegraphics[width=150mm]{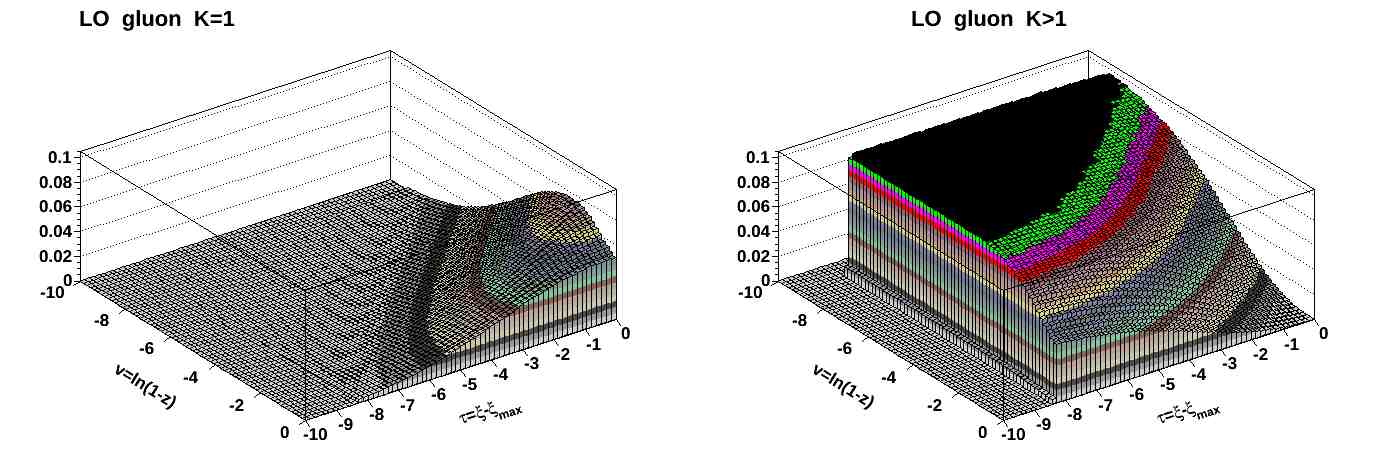}}
  \caption{
    The inclusive LO distribution of
    Fig.~\protect\ref{fig:mcRho1glu}
    split into the hardest gluon (left) and the rest (right).
    }
  \label{fig:mcCanv1k}
\end{figure*}

Figure~\ref{fig:mcCanv1k} demonstrates a split of the LO inclusive
distribution of Fig.~\ref{fig:mcRho1glu}
into the $K=1$ component and the rest $K>1$.
The important point is that the $K=1$ component
reproduces the original complete distribution over the whole region
where the NLO correction is non-negligible!
This is exactly the observation on which POWHEG technique is built.
According to the POWHEG authors, taking the $K=1$ component is
sufficient to reproduce the complete NLO correction (modulo NNLO).

\begin{figure}[!ht]
  \centering
  {\includegraphics[width=0.8\textwidth,height=35mm]{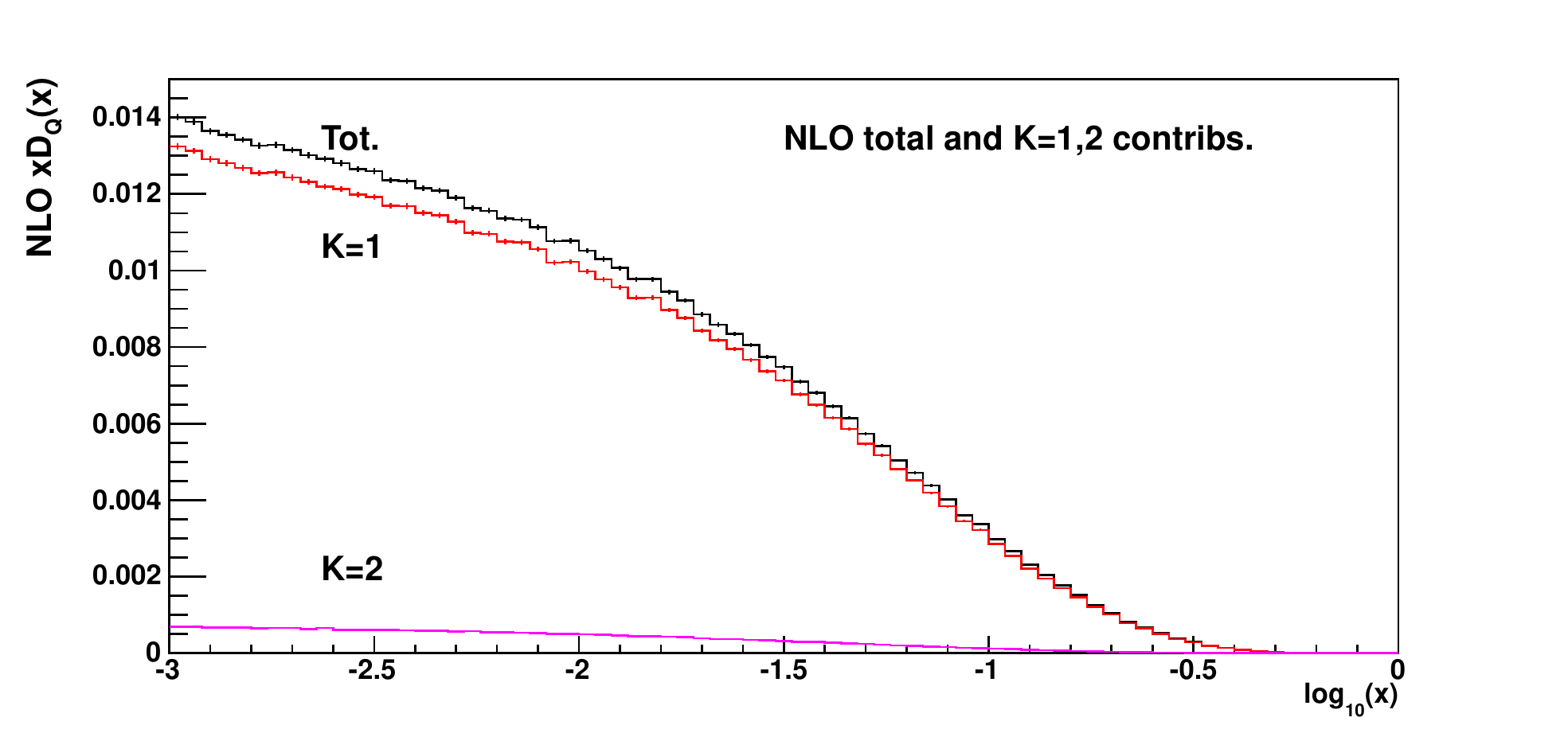}}
  \caption{
    The original NLO correction from $\sum_j W^{NLO}_j$ 
    and its two hardest gluon
    components, from $W^{NLO}_{K=1}$ and $W^{NLO}_{K=2}$,
    as a function of $x=\prod_j z_j$.
    }
  \label{fig:mcCanV8k}
\end{figure}

The above statement is checked numerically
in Fig.~\ref{fig:mcCanV8k},
where we compare the NLO correction
to the $x=\prod_j z_j$ distribution from the complete sum
$\sum_j W^{NLO}_j$ and from $ W^{NLO}_{K=1}$.
As we see the $K=1$ component saturates the complete sum very well,
with the $K=2$ component being negligible in the first approximation.

We can therefore speed up the calculation
by means of taking only the $K=1$ contribution.
The price will be that 
the formula of eq.~(\ref{eq:DYanxch}) will
not be exact any more.
Our method differs, however, from the POWHEG
scheme, where the $K=1$ gluon is generated
separately in the first step,
and other gluons are generated (by the LO parton shower MC)
in the next step.
That is easy for LO MC with $k^T$-ordering, while in case
of the LO MC with angular-ordering POWHEG requires additional effort of
generating the so called vetoed and truncated showers.
In our method,
there is no need for such vetoed/truncated showers
in case of angular ordering.

\begin{figure*}[!ht]
  \centering
  {\includegraphics[width=140mm,height=80mm]{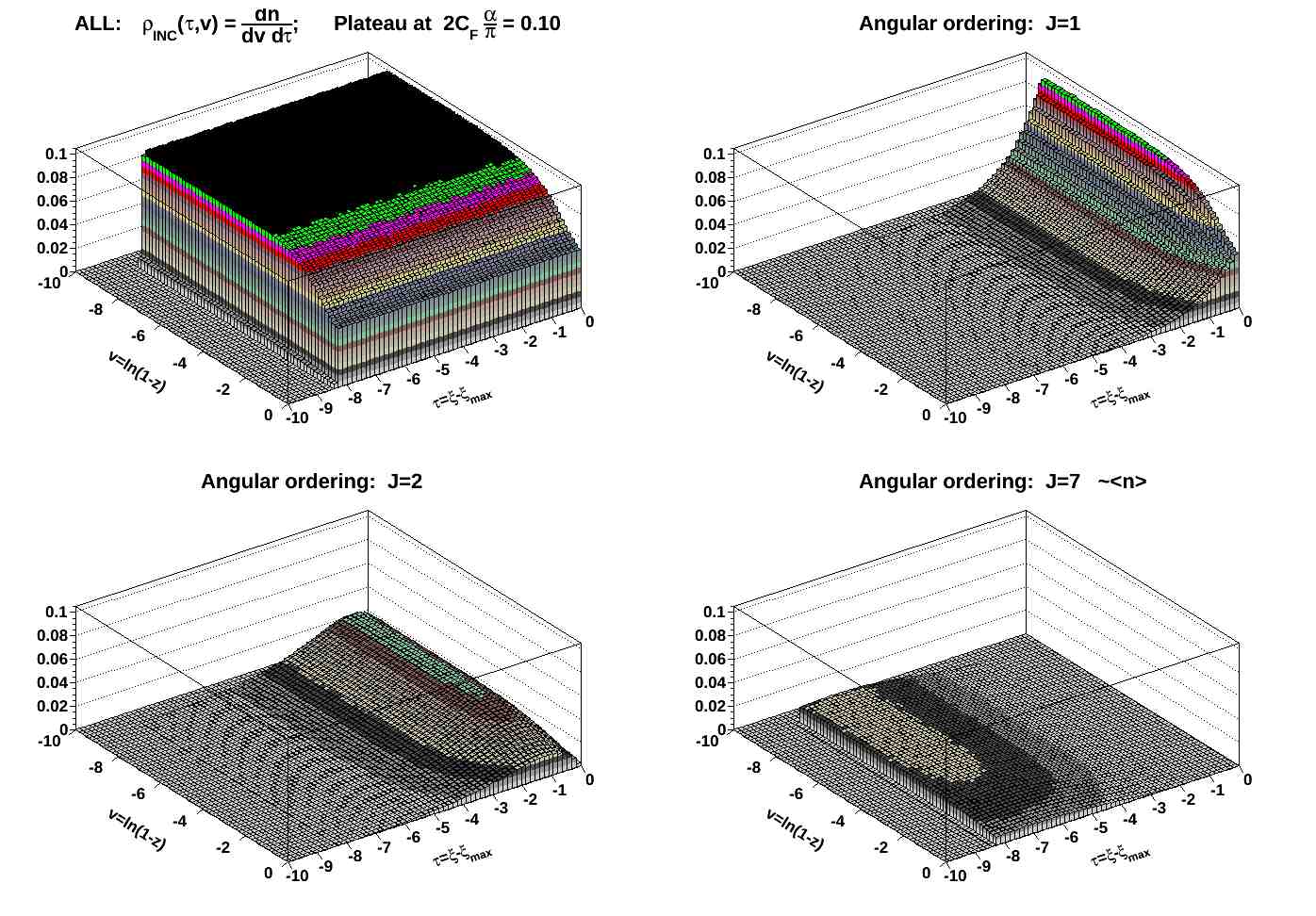}}
  \caption{
    The distribution of gluons ordered in rapidity,
    as in our basic LO MC.
    }
  \label{fig:mcRhoOrdLO}
\end{figure*}

The reason why POWHEG technique is complicated in case
of the angular ordering is illustrated in Fig.~\ref{fig:mcRhoOrdLO}. 
We show there the distribution of gluons ordered in rapidity,
starting from the gluon with the maximum rapidity, 
the closest to hard process.
The gluon distribution
with the highest rapidity $\xi\sim \xi_{\max}$ ($J=1$)
has a ridge extending towards the soft region.
Notice that, when the IR cut-off $\epsilon\to0$ in $(1-z)<\epsilon$,
the width of this ridge also goes to zero.
Consequently, the gluon with the highest $\xi$ is unable
to reproduce the gluon distribution in the NLO corner,
close to hard process.
This is why in this case POWHEG requires truncated and vetoed showers,
which are not needed in our method.

\section{Summary and outlook}
A new method of adding 
the QCD NLO corrections to the hard process
in the initial state Monte Carlo parton shower
is tested numerically showing
that the basic concept of the new methodology
works correctly in the numerical environment of 
a Monte Carlo parton shower.
The differences with the well established methods
of MC@NLO and POWHEG are briefly discussed.
Also, variants of the new method with
better efficiency in terms of CPU time are proposed.

\section*{Acknowledgement}
This work is partly supported by 
the Polish National Science Centre grant UMO-2012/04/M/ST2/00240,
Foundation for Polish Science grant Homing Plus/2010-2/6,
  the Research Executive Agency (REA) of the European Union 
  Grant PITN-GA-2010-264564 (LHCPhenoNet),
the U.S.\ Department of Energy
under grant DE-FG02-04ER41299 and the Lightner-Sams Foundation.


\end{document}